\documentclass[aps,prd,manuscript,amsfonts]{revtex4}

\begin{document}

\title{The Schr\" odinger picture of the Dirac quantum mechanics on spatially
flat Robertson-Walker backgrounds}

\author{Ion I.  Cot\u aescu \thanks{E-mail:~~~cota@physics.uvt.ro}\\
{\small \it West University of Timi\c soara,}\\
       {\small \it V.  P\^ arvan Ave.  4, RO-300223 Timi\c soara, Romania}}


\begin{abstract}
The Schr\" odinger picture of the Dirac quantum mechanics is defined in charts
with spatially flat Robertson-Walker metrics and Cartesian coordinates. The
main observables of this picture are identified, including the interacting part
of the Hamiltonian operator produced by the minimal coupling with the
gravitational field. It is shown that in this approach new Dirac quantum modes
on de Sitter spacetimes may be found analytically solving the Dirac equation.

 Pacs:
04.62.+v
\end{abstract}

\maketitle

The relativistic quantum mechanics of the spin-half particle on a given
background can be constructed as the one-particle restriction of the quantum
theory of the free Dirac field on this background, considered as a perturbation
that does not affect the geometry. The central piece is the Dirac equation
whose form depends on the local chart (or natural frame) and the tetrad fields
defining the local frames and co-frames. This type of quantum mechanics has two
virtues. First of all the charge conjugation of the Dirac field is
point-independent indicating that the vacuum of the original field theory is
stable in any geometry \cite{co,cot}. Therefore, the resulted one-particle
Dirac quantum mechanics can be seen as a coherent theory similar to that of
special relativity. The second virtue is just the spin which generates specific
terms helping us to correctly interpret the physical meaning of principal
operators.

In the non-relativistic quantum mechanics the time evolution can be studied in
different pictures (e. g., Schr\" odinger, Heisenberg, Interaction) which
transform among themselves through  specific time-dependent unitary
transformations. It is known that the form of the Hamiltonian operator and the
time dependence of other operators strongly depend on the picture choice. In
special and general relativity, despite of its importance, the problem of
time-evolution pictures is less studied because of the difficulties in finding
suitable Hamiltonian operators for scalar or vector fields. However, the Dirac
quantum mechanics is a convenient framework for studying this problem since the
Dirac equation can be put in Hamiltonian form at any time.

In this paper we should like to show that at least two different pictures of
the Dirac quantum mechanics can be identified in the case of backgrounds with
spatially flat Robertson-Walker (RW) metrics. We start with the simple
conjecture of the Dirac equation in diagonal gauge and Cartesian coordinates
considering that this constitutes the {\em natural} picture. Furthermore, we
define the Schr\" odinger picture such that the kinetic part of the Dirac
equation should take the standard form known from special relativity. In this
picture we identify the momentum and the Hamiltonian operators pointing out
that they represent a generalization of the similar operators we obtained
previously on de Sitter spacetimes \cite{cot}.

Let us start denoting by $\{t,\vec{x}\}$ the Cartesian coordinates $x^{\mu}$
($\mu,\nu,...=0,1,2,3 $)  of a chart with the RW line element
\begin{equation}
ds^2=g_{\mu\nu}(x)dx^{\mu}dx^{\nu}=dt^2-\alpha(t)^2 (d\vec{x}\cdot d\vec{x})
\end{equation}
where $\alpha$ is an arbitrary time dependent function. In this chart we
introduce the tetrad fields $e_{\hat\mu}(x)$ that define the local frames  and
those defining the corresponding coframes, $\hat e^{\hat\mu}(x)$ \cite{SW}.
These fields are labeled by the local indices ($\hat\mu,\hat\nu,...=0,1,2,3$)
of the Minkowski metric $\eta=$diag$(1,-1,-1,-1)$, satisfy $e_{\hat\mu}(x)\hat
e^{\hat\mu}(x)=1_{4\times4}$ and give the metric tensor as $g_{\mu
\nu}=\eta_{\hat\alpha\hat\beta}\hat e^{\hat\alpha}_{\mu}\hat
e^{\hat\beta}_{\nu}$. Here we consider the tetrad fields of the diagonal gauge
that have non-vanishing components \cite{BD,SHI},
\begin{equation}\label{tt}
e^{0}_{0}=1\,, \quad e^{i}_{j}=\frac{1}{\alpha(t)}\delta^{i}_{j}\,,\quad \hat
e^{0}_{0}=1\,, \quad \hat e^{i}_{j}=\alpha(t)\delta^{i}_{j}\,,\quad
i,j,...=1,2,3\,,
\end{equation}
determining the form of the Dirac equation \cite{BD},
\begin{equation}\label{ED1}
\left(i\gamma^0\partial_{t}+i\frac{1}{\alpha(t)}\gamma^i\partial_i
+\frac{3i}{2}\frac{\dot{\alpha}(t)}{\alpha(t)}\gamma^{0}-m\right)\psi(x)=0\,.
\end{equation}
This is expressed in terms of Dirac $\gamma$-matrices \cite{TH} and the fermion
mass $m$, with the notation $\dot{\alpha}(t)=\partial_t\alpha(t)$. Thus we
obtain the natural picture in which the time evolution is governed by the Dirac
equation (\ref{ED1}). The principal operators of this picture, the energy $\hat
H$, momentum $\vec{\hat P}$ and coordinate $\vec{\hat X}$, can be defined as in
special relativity,
\begin{equation}\label{ON}
(\hat H \psi)(x)=i\partial_t\psi(x)\,,\quad (\hat P^i
\psi)(x)=-i\partial_i\psi_S(x)\,,\quad (\hat X^i \psi)(x)=x^i\psi(x)\,.
\end{equation}
The operators $\hat X^i$ and $\hat P^i$ are time-independent and satisfy the
well-known canonical commutation relations
\begin{equation}\label{com}
\left[\hat X^i, \hat P^j\right]=i\delta_{ij}I\,,\quad \left[\hat H, \hat
X^i\right]=\left[\hat H,\hat P^i\right]=0\,,
\end{equation}
where $I$ is the identity operator. Other operators are formed by orbital parts
and suitable spin parts that can be point-dependent too. In general, the
orbital terms are freely generated by the basic orbital operators $\hat X^i$
and $\hat P^i$. An example is the total angular momentum
$\vec{J}=\vec{L}+\vec{S}$ where $\vec{L}=\vec{\hat X}\times\vec{\hat P}$ and
$\vec{S}$ is the spin operator. We specify that the operators $\hat P^i$ and
$J^i$ are generators of the spinor representation of the isometry group $E(3)$
of the spatially flat RW manifolds \cite{cot}. Therefore, these operators are
{\em conserved} in the sense that they commute with the Dirac operator
\cite{CML,ES}.

The natural picture can be changed using point-dependent operators
which could be even non-unitary operators since the relativistic
scalar product does not have a direct physical meaning as that of
the non-relativistic quantum mechanics. We exploit this opportunity
for defining the Schr\" odinger picture as the picture in which the
kinetic part of the Dirac operator takes the standard form
$i\gamma^0\partial_t+i\gamma^i\partial_i$. The transformation
$\psi(x)\to \psi_S(x)=U_S(x)\psi(x)$ leading to the Schr\" odinger
picture is produced by the operator of time dependent {\em
dilatations}
\begin{equation}\label{U}
U_S(x)=\exp\left[-\ln(\alpha(t))(\vec{x}\cdot\vec{\partial})\right]\,,
\end{equation}
 which has the following suitable action
\begin{equation}
U_S(x)F(\vec{x})U_S(x)^{-1}=F\left(\frac{1}{\alpha(t)}\vec{x}\right)\,,\quad
U_S(x)G(\vec{\partial})U_S(x)^{-1}=G\left(\alpha(t)\vec{\partial}\right)\,,
\end{equation}
upon any analytical functions $F$ and $G$. Performing this transformation we
obtain the Dirac equation of the Schr\" odinger picture
\begin{equation}\label{ED2}
\left[i\gamma^0\partial_{t}+i\vec{\gamma}\cdot\vec{\partial} -m
+i\gamma^{0}\frac{\dot{\alpha}(t)}{\alpha(t)}
\left(\vec{x}\cdot\vec{\partial}+\frac{3}{2}\right)\right]\psi_S(x)=0\,.
\end{equation}
Hereby we have to identify the specific operators of this picture, the energy
$H_S$ and the operators $P^i_S$ and  $X^i_S$ that must be time-independent, as
in the non-relativistic case. We assume that these operators are defined as
\begin{equation}\label{OS}
(H_S \psi_S)(x)=i\partial_t\psi_S(x)\,,\quad (P^i_S
\psi_S)(x)=-i\partial_i\psi_S(x)\,,\quad (X^i_S \psi_S)(x)=x^i\psi_S(x)\,,
\end{equation}
obeying commutation relations similar to Eqs. (\ref{com}). The Dirac equation
(\ref{ED2}) can be put in Hamiltonian form, $H_S\psi_S={\cal H}_S\psi_S$, where
the Hamiltonian operator ${\cal H}_S={\cal H}_0 + {\cal H}_{int}$ has the
standard kinetic term ${\cal H}_0=\gamma^0\vec{\gamma}\cdot \vec{P}_S+\gamma^0
m$ and the interaction term with the gravitational field,
\begin{equation}\label{Hint}
{\cal H}_{int}=\frac{\dot{\alpha}(t)}{\alpha(t)}\left(\vec{X}_S\cdot
\vec{P}_S-\frac{3i}{2}I\right)=\frac{\dot{\alpha}(t)}{\alpha(t)}\left(\vec{\hat
X}\cdot \vec{\hat P}-\frac{3i}{2}I\right)\,,
\end{equation}
which vanishes in the absence of gravitation when $\alpha$ reduces to a
constant.

The sets of operators (\ref{OS}) and (\ref{ON}) are defined in different
manners such that they have similar expressions but in different pictures. For
analyzing the relations among these operators it is convenient to turn back to
the natural picture. Performing the inverse transformation we find that in this
picture the operators (\ref{OS}) become new interesting time-dependent
operators,
\begin{eqnarray}
H(t)&=&U_S(x)^{-1}H_SU_S(x)=\hat H+\frac{\dot{\alpha}(t)}{\alpha(t)}
\vec{\hat X}\cdot\vec{\hat P}\,,\label{Ht}\\
X^i(t)&=&U_S(x)^{-1}X_S^iU_S(x)=\alpha(t) \hat X^i\,,\\
P^i(t)&=&U_S(x)^{-1}P_S^iU_S(x)=\frac{1}{\alpha(t)}\hat P^i\,,\label{Pt}
\end{eqnarray}
satisfying the usual commutation relations (\ref{com}). Notice that the total
angular momentum and the operator (\ref{Hint}) have the same expressions in
both these pictures since they commute with $U_S(x)$.

Now the problem is to select the set of operators with a good physical meaning.
We observe that in the moving charts with RW metrics of the de Sitter spacetime
the operator (\ref{Ht}) is time-independent (since $\dot{\alpha}/\alpha
=$const.) and {\em conserved}, corresponding to the unique time-like Killing
vector of the $SO(4,1)$ isometries \cite{cot}. This is an argument indicating
that the correct physical observables are the operators (\ref{Ht})-(\ref{Pt})
while the operators (\ref{ON}) may be considered as auxiliary ones. In fact
these are just the usual operators of the relativistic quantum mechanics on
Minkowski spacetime where is no gravitation. The examples we worked out
\cite{cot,C,C1} convinced us that this is the most plausible interpretation
even though this is not in accordance with other attempts  \cite{Guo}.

The Schr\" odinger picture we defined above may offer one some technical
advantages in solving problems of quantum systems interacting with the
gravitational field. For example, in this picture we can derive the
non-relativistic limit (in the sense of special relativity) replacing ${\cal
H}_0$ directly by the Schr\" odinger kinetic term $\frac{1}{2m}{\vec{P}_S}^2$.
Thus we obtain the Schr\" odinger equation
\begin{equation}\label{Sc}
\left[-\frac{1}{2m}\Delta -i\,
\frac{\dot{\alpha}(t)}{\alpha(t)}\left(\vec{x}\cdot \vec{\partial
}+\frac{3}{2}\right)\right]\phi(x)=i\partial_t \phi(x)\,,
\end{equation}
for the wave-function $\phi$ of a spinless particle of mass $m$. Moreover,
using standard methods one can derive the next approximations in $1/c$
producing characteristic spin terms. It is remarkable that the non-relativistic
Hamiltonian operators obtained in this way are Hermitian with respect to the
usual non-relativistic scalar product.

In the particular case of the de Sitter spacetime, the  Schr\" odinger picture
will lead to important new results for the Dirac and Schr\" odinger equations
in moving charts with RW metrics and spherical coordinates. In these charts
where $H=H(t)$ is conserved both the mentioned equations, namely Eqs.
(\ref{ED2}) and (\ref{Sc}), are analytically solvable in terms of Gauss
hypergeometric functions and, respectively, Whittaker ones \cite{coco}.
Therefore, in this picture it appears the opportunity of deriving new Dirac
quantum modes determined by the set of commuting operators $\{H, {\vec{J}\,}^2,
J_3, K\}$ where $K=\gamma^0(2\vec{L}\cdot\vec{S}+1)$ is the Dirac angular
operator. We specify that common eigenspinors of this set of operators were
written down in {\em static} central charts with spherical coordinates \cite{C}
but never in moving charts. We remind the reader that in moving charts  with
spherical coordinates one knows only the Shishkin's solutions  of the Dirac
equation \cite{SHI} derived in the natural picture. We have shown \cite{C1}
that there are suitable linear combinations of these solutions representing
common eigenspinors of the set $\{{\vec{\hat P}\,}^2,{\vec{J}\,}^2, J_3, K\}$.
Taking into account that the operators $H$ and ${\vec{\hat P}\,}^2$ do not
commute with each other we understand the importance of the new quantum modes
that could be showed off grace to our Schr\" odinger picture.

Finally we note that the quantum mechanics developed here is a specific
approach working only in spatially flat RW geometries. This may be completed
with new pictures (as the Heisenberg one) after we shall find the general
mechanisms of time evolution in the relativistic quantum mechanics. However,
now it is premature to look for general principles  before to carefully analyze
other significant particular examples.

\subsection*{Acknowledgments}

We are grateful to Erhardt Papp for interesting and useful discussions  on
closely related subjects.


\begin{thebibliography}{20}

\bibitem{co}
I. I. Cot\u aescu, {\em Int. J. Mod. Phys. A} {\bf 19}, 2117 (2004).

\bibitem{cot}
I. I. Cot\u aescu, {\em Phys. Rev. D} {\bf 65}, 084008 (2002).

\bibitem{SW}
S. Weinberg, {\it Gravitation and Cosmology: Principles and Applications of the
General Theory of Relativity}  (Wiley, New York, 1972).

\bibitem{BD}
A. O. Barut and I. H. Duru, {\em Phys. Rev. D} {\bf 36}, 3705 (1987).

\bibitem{SHI}
G. V. Shishkin, {\it Class. Quantum Grav.} {\bf 8}, 175 (1991).

\bibitem{TH}
B. Thaller,  {\it The Dirac Equation}, Springer Verlag, Berlin Heidelberg, 1992

\bibitem{CML}
B. Carter and R. G. McLenaghan, {\em Phys. Rev. D} {\bf 19}, 1093 (1979).

\bibitem{ES}
I. I. Cot\u aescu, {\em J. Phys. A: Math. Gen.} {\bf 33}, 1977
(2000).

\bibitem{C}
I. I. Cot\u aescu, {\em Mod. Phys. Lett. A} {\bf 13}, 2991  (1998).

\bibitem{C1}
I. I. Cot\u aescu, Radu Racoceanu and Cosmin Crucean, Mod. Phys. Lett. A {\bf
21}, 1313 (2006).

\bibitem{Guo}
 H.-Y. Guo, C.-G. Huang, Z. Xu and B. Zhou, {\em Mod. Phys. Lett. A} {\bf 19},
 1701 (2004);  id. {\tt hep-th/0311156}; E. A. Tigrov, {\tt gr-qc/0011011}.

\bibitem{coco}
I. I. Cot\u aescu, in preparation.




\end{thebibliography}
\end{document}